\documentclass[aps,pre,twocolumn,showpacs,floatfix]{revtex4}
\usepackage{graphicx,bm,amssymb,amsmath,dcolumn,}
\usepackage{epstopdf}   
\usepackage{color}
\begin{document}
\newcommand{\be}{\begin{equation}}
\newcommand{\ee}{\end{equation}}
\newcommand{\half}{\frac{1}{2}}
\newcommand{\ith}{^{(i)}}
\newcommand{\im}{^{(i-1)}}
\newcommand{\gae}
{\,\hbox{\lower0.5ex\hbox{$\sim$}\llap{\raise0.5ex\hbox{$>$}}}\,}
\newcommand{\lae}
{\,\hbox{\lower0.5ex\hbox{$\sim$}\llap{\raise0.5ex\hbox{$<$}}}\,}


\definecolor{blue}{rgb}{0,0,1}
\definecolor{red}{rgb}{1,0,0}
\definecolor{green}{rgb}{0,1,0}
\newcommand{\blue}[1]{\textcolor{blue}{#1}}
\newcommand{\red}[1]{\textcolor{red}{#1}}
\newcommand{\green}[1]{\textcolor{green}{#1}}

\newcommand{\scrA}{{\mathcal A}} 
\newcommand{\scrN}{{\mathcal N}} 
\newcommand{\scrS}{{\mathcal S}} 
\newcommand{\scrP}{{\mathcal P}} 
\newcommand{\scrO}{{\mathcal O}} 
\newcommand{\scrR}{{\mathcal R}} 

\newcommand{\dm}{d_{\rm min}} 

\title{Shortest-Path Fractal Dimension for Percolation in Two and Three Dimensions}
\author{Zongzheng Zhou$^{1}$, Ji Yang$^{1}$,  
Youjin Deng$^1\footnote{Email: yjdeng@ustc.edu.cn}$, 
Robert M. Ziff$^2\footnote{Email: rziff@umich.edu}$} 
\affiliation{$^{1}$Hefei National Laboratory for Physical Sciences
at the Microscale and Department of Modern Physics,
University of Science and Technology of China, Hefei, Anhui 230027, PR China }
\affiliation{{$^2$}Michigan Center for Theoretical Physics and Department of Chemical
Engineering, University of Michigan, Ann Arbor, Michigan 48109-2136, USA}

\date{\today} 
\begin{abstract}
 We carry out a high-precision Monte Carlo study of the shortest-path fractal dimension $\dm$
 for percolation in two and three dimensions, using the Leath-Alexandrowicz method
 which grows a cluster from an active seed site. 
 A variety of quantities are sampled as a function of the chemical distance, including the number of activated sites, 
 a measure of the radius, and the survival probability.
 By finite-size scaling, we determine  $\dm = 1.130\,77(2)$ and  $1.375\,6(6)$ in two and three dimensions, respectively. 
 The result in two dimensions rules out the recently conjectured value $\dm=217/192$ [Phys.\ Rev.\ E {\bf 81}, 020102(R) (2010)].
\end{abstract}
\maketitle 

\section{Introduction}
 As a standard model of disordered systems \cite{Stauffer-1994,Bunde-1996},
 percolation has been intensively studied over the last 50 years and applied to many other fields
 due to its richness in both mathematics and physics.
 The nature of the phase transitions in percolation has been well established.
 In particular, within the two-dimensional (2D) universality class,
 there are only a few critical exponents left to be expressed exactly,
 among which is the shortest-path fractal dimension $\dm$, defined by
 \cite{Bunde-1996,Grassberger-92-0,Grassberger-92-1}
 \begin{equation}
 \langle \ell \rangle \sim r^{\dm},
 \end{equation}
 where $r$ is the Euclidean distance between two sites belonging to the same
 cluster, and $\ell$ is the shortest path.

 The shortest path $\ell$ between two sites in a cluster is
 the minimum number of steps on a path of occupied bonds or sites in the cluster,
 and was first studied independently by several groups in the early 1980s
 \cite{Grassberger-83,Alexandrowicz-80,Pike-81,Middlemiss-80,Herrmann-84}.
 The length $\ell$ is also called the chemical distance \cite{HavlinNossal84}.
 A related quantity is the spreading dimension $d_{\ell}$ \cite{VannimenusNadalMartin84},
 which describes the scaling of the mass $\scrN$ of a critical cluster within 
 a chemical distance $\ell$ as $\scrN \sim \ell^{d_\ell}$, and is related to 
 the fractal dimension $d_f$ of the cluster by $d_\ell = d_f/\dm$.

 In percolation, the shortest-path naturally occurs during epidemic growth
 or burning algorithms.
 Previous measurements of $\dm$ in 2D include $\dm=1.18(4)$ \cite{Alexandrowicz-80}, 1.118(15) \cite{Pike-81},
 1.15(3) \cite{HavlinNossal84}, 1.102(13) \cite{RammalEtAl84}, 1.132(4) \cite{Grassberger85},
 1.130(2) \cite{HerrmannStanley88}, 1.130\,6(3) \cite{Grassberger99} and 1.1303(8) \cite{DengEtAl10}.
 A summary of the early work is given in Ref.\ \cite{Grassberger85}.
 
 In 1984, Havlin and Nossal \cite{HavlinNossal84} conjectured that
 $\dm = d_f - 1/\nu = 91/48-3/4=55/48 = 1.145\,833$,
 which was soon shown to be too large \cite{HerrmannStanley88,Grassberger85}.
 In 1987 Larsson \cite{Larsson} speculated that $\dm$ could be 17/16 or even 1,
 but these are both excluded.
 In 1988 Herrmann and Stanley \cite{HerrmannStanley88} conjectured that $\dm = 2 - d_B +d_{\rm red}$,
 where $d_{\rm red} = 1/\nu = 3/4$ is the ``red''-bond dimension and
 $d_B$ is the backbone dimension.
 Using Deng, Bl\"{o}te and Neinhuis's result $d_B=1.643\,4(2)$ \cite{DengBloteNienhuis04}
 (see also \cite{Grassberger99b,JacobsenZinnJustin02}),
 we find that this prediction gives $\dm =2-1.643\,4(2)+0.75=1.106\,6(2)$, which is too small.
 In 1989, Tzschichholz, Bunde and Havlin \cite{TzschichholzBundeHavlin89} considered
 $\dm = 53/48=1.104\,166\,6...$, which is also below measured values.

 In 1998, Porto {\it et al}.~\cite{Porto} conjectured that $\dm$ is related to
 a pair-connectivity scaling exponent $g_1$ by $\dm = g_1 + \beta/\nu$ where $\beta=5/36$ for 2D.
 However, $g_1$ was later shown to have the exact value $g_1 = 25/24$ \cite{Ziff99,Grassberger99}, 
 which implies $\dm = 55/48 = 1.145\,833$, identical to Havlin and Nossal's earlier conjecture \cite{HavlinNossal84}.

 In 2010, one of us (Y.D.) and coauthors
 conjectured an exact expression \cite{DengEtAl10} of $\dm$ for 
 the 2D critical and tricritical random-cluster model: $\dm = (g + 2)/(g + 18)/32g$, where $g$ is
 the Coulomb-gas coupling constant, related to the random cluster fugacity $q$ by $g = (2/\pi)\cos^{-1}(q/2-1)$.
 This conjecture is numerically correct up to the third or fourth decimal place for all values of $q$ 
 studied in Ref.\ \cite{DengEtAl10}.
 For the $q \rightarrow 1$ limit---i.\,e., standard bond percolation---the predicted value $\dm=217/192=1.130\,208$ was consistent with the numerical results 
 in previous works \cite{Alexandrowicz-80,Pike-81,HavlinNossal84,RammalEtAl84,Grassberger85,HerrmannStanley88,Grassberger99}.
 In addition, the conjectured formula exhibits other good properties.
 It reproduces the exact results for the critical uniform spanning tree ($q \rightarrow 0$)
 as well as for the tricritical $q \rightarrow 0$ Potts model;
 at the tricritical $q \rightarrow 0$ point, the derivative with respect to $q$ is also correct.

 The main goal of the present work is to carry out a high-precision Monte Carlo test of 
 the conjecture in Ref.\ \cite{DengEtAl10} in the context of 2D percolation.
 A numerical estimate of $\dm$ for 3D percolation is also provided.
 Some preliminary results of this work were reported in a
 recent paper on biased directed percolation \cite{Zhou12}.

\section{simulation and sampled quantities}
 We simulate bond percolation on the square and the simple-cubic lattice by the Leath-Alexandrowicz 
 algorithm \cite{Leath,Alexandrowicz-80}, which grows a percolation cluster starting from a seed site.
 For each neighboring edge of the seed site an occupied bond is placed with occupation probability $p$,
 and the neighboring site is activated and added into the growing cluster.
 After all the neighboring edges of the seed site have been visited, the growing procedure is continued
 from the newly added sites. This proceeds until no more new sites can be added into the cluster (the procedure dies out) 
 or the initially set maximum time step $\ell_{\rm max}$ is reached.

 The above procedure is also called breadth-first growth, and $\ell$ is equal to the shortest-path length 
 between the seed site and any activated sites at time step $\ell$. 
 We set $\ell=1$ for the beginning of the growth, and measure the number of activated sites $N(\ell)$ as a function of $\ell$.
 In addition, we record the Euclidean distance $r_i$ of each activated site $i$ to the seed site, and
 define a radius by
 \be
 \label{eq:gyration_radius}
 R(\ell) = \left\{ \begin{array} {ll}
             0                              & \hspace{10mm} \mbox{if } N(\ell) = 0     \\
	     \sqrt{\sum_{i=1}^N r_i^2}      & \hspace{10mm} \mbox{if } N(\ell) \geq 1  \ . \\
	     \end{array} 
	     \right.
 \ee
 The statistical averages, $\scrN(\ell) \equiv \langle N(\ell) \rangle$ and $\scrR(\ell) \equiv {\langle R^2(\ell) \rangle}^{1/2}$,
 and the associated error bars are calculated~\footnote{The definition of $\scrR$ here is different than that of Ref.~\cite{Zhou12}, where
simply the average of $R$ was used. We found that by using the average square radius, a better estimate of $d_{\rm min}$ was obtained.}.
 We also sample the survival probability $\scrP(\ell)$ that at time step $\ell$, the growing procedure still survives.

 At criticality, one expects the scaling behavior 
 \begin{equation}
  \scrN(\ell) \sim  \ell^{ Y_N} \; , \hspace{5mm} \scrR(\ell) \sim  \ell^{ Y_R} \; , \hspace{5mm}   \scrP(\ell) \sim  \ell^{-Y_P} \; ,  
  \label{eq:scaling}
  \end{equation}
 where the critical exponents $Y_P$, $Y_N$, and $Y_R$ are related to $\beta$, $\nu$ and $\dm$ by
 \begin{eqnarray}
   Y_N &=& \frac{\gamma}{\nu \dm} - 1 \, ,  \hspace{5mm}   Y_P = \frac{\beta}{\nu \dm} \, , \nonumber \\
   2Y_R &=& \frac{\gamma + 2\nu}{\nu \dm} - 1\; , 
 \label{eq:scaling_relation}
 \end{eqnarray}
 with $\gamma = d\nu - 2\beta$ and $d$ equal to the spatial dimensionality.
 In terms of exponents of epidemic processes \cite{GrassbergerDeLaTorre},
 these quantities correspond to $\delta = Y_P$, $\eta = Y_N$, and $1/z - \eta/2 = Y_R$.

 To eliminate the unknown non-universal constants in front of the scaling behaviors (\ref{eq:scaling}), 
 we define the ratio $Q_\scrO (\ell) = \scrO (2 \ell)/\scrO (\ell)$ for $\scrO=\scrN, \scrR$ and $\scrP$. 
 In the $\ell \rightarrow \infty$ limit, one has
 \begin{equation}
   Y_N = \log_2 (Q_\scrN) , \  
   Y_R = \log_2 (Q_\scrR) , \  
   Y_P = -\log_2 (Q_\scrP) . 
   \label{eq:exponent-Q}
 \end{equation}

 In 2D, one has the exactly known exponents $\beta = 5/36$, $\nu = 4/3$, and $\gamma = 43/18$ \cite{Stauffer-1994,Cardy,Smirnov,Lawler,Kesten-1987}. 
 In 3D, the exact values are unknown, and are numerically found 
 to be $\beta/\nu = 0.4774(1)$ and $\nu =0.8764(7)$~\cite{Deng-05,Martins-03-PRE,Tomita-JPS-02,
 Ballesteros-99-JPA,Junfeng}.   

\section{Initial estimate of $\dm$ for 2D}

 We first carried out simulations at the critical point $p=1/2$ for bond percolation 
 on the square lattice with time step up to $\ell_{\rm max} = 1024$ and the number of samples about  $2\times 10^9$.
  
 The asymptotic behavior of the observables $\scrN$, $\scrR$ and $\scrP$ is expected to follow the form
 \begin{equation}
 \label{eq:Pt}
 \scrO(\ell) = \ell^{Y} (a_0 + b_1\ell^{y_1} + b_2\ell^{-2} ) \; ,
 \end{equation}
 where higher-order corrections are neglected and the critical exponent $Y$ is given by Eq.\ (\ref{eq:scaling}).
 The leading finite-size correction exponent is known to be $y_1 =-0.96 (6) \approx -1$ \cite{Zhou12}.
A least-squares criterion was used to fit the data assuming the above form.
 With $y_1$ being fixed at $-1$, the fit of $\scrP$ gives 
 $\dm = 1.1308 \pm 0.0002$ and $b_1 = 0.045 (5)$, and the fit of $\scrN$ yields $\dm = 1.1308\pm 0.0002$.

\begin{figure}
 \includegraphics[scale=0.38]{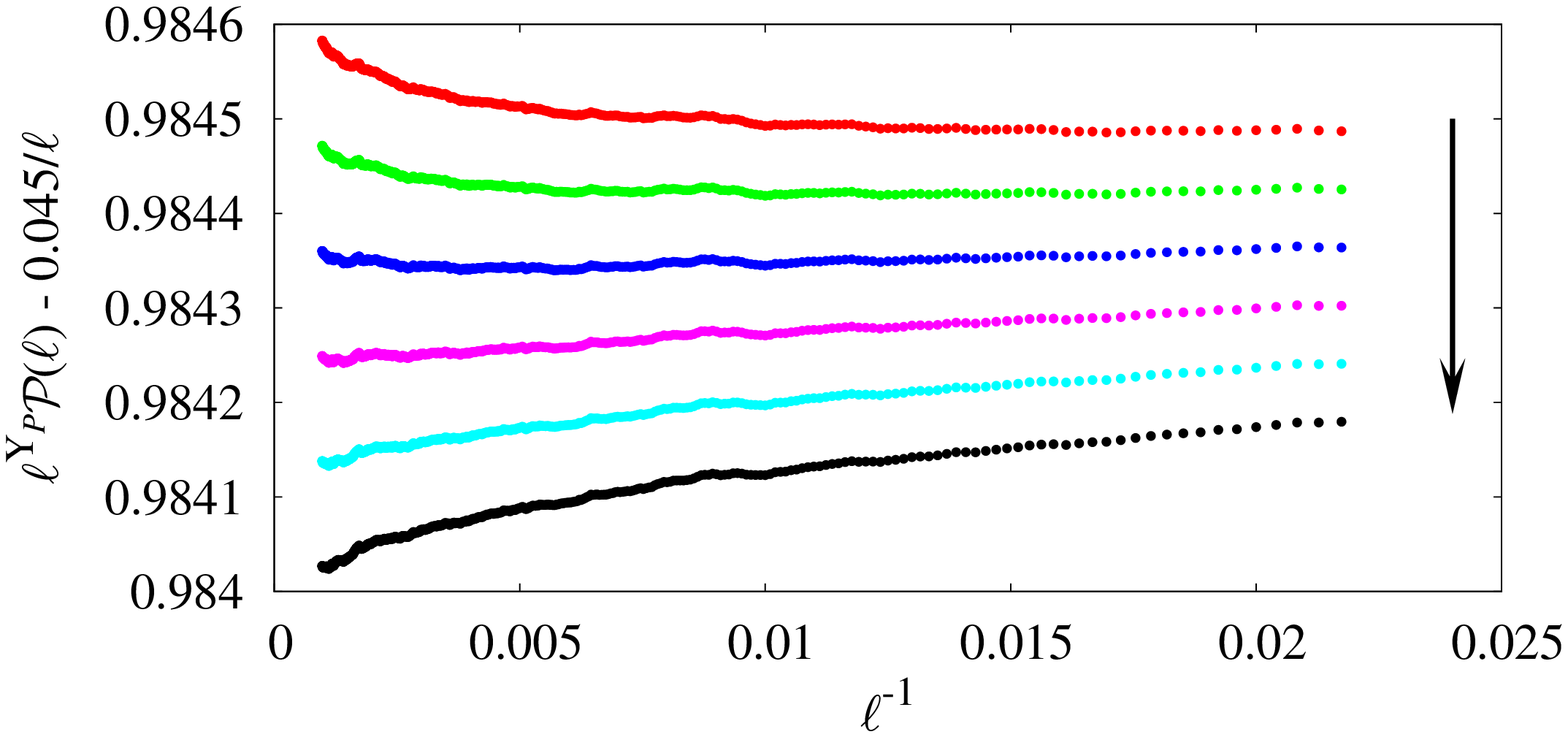}
 \centering
 \caption{(Color online) Plot of $\scrP \ell^{Y_P} - 0.045\ell^{-1}$ versus $\ell^{-1}$ in 2D. 
          The $Y_P$ value is obtained via Eq.\ (\ref{eq:scaling_relation}) by setting $\dm$ 
	  at 1.1302, 1.1304, 1.1306, 1.1308, 1.1310 and 1.1312, following the arrow.}
 \label{fig:P}
 \end{figure}

\begin{figure}
 \includegraphics[scale=0.65]{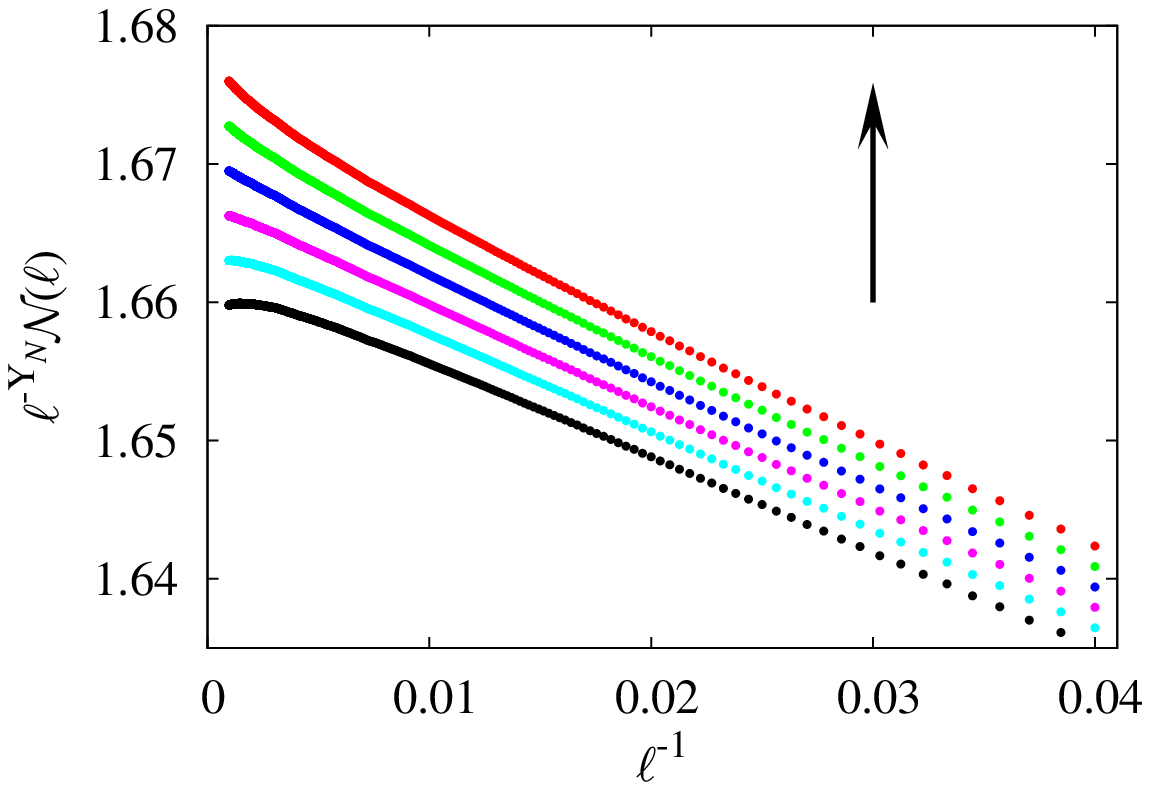}
 \centering
 \caption{(Color online) Plot of $\scrN \ell^{-Y_N}$ versus $\ell^{-1}$ in 2D. 
          The $Y_N$ value is obtained via Eq.\ (\ref{eq:scaling_relation}) by setting $d_{\rm min}$ 
	  at 1.1302, 1.1304, 1.1306, 1.1308, 1.1310 and 1.1312, following the arrow.}
 \label{fig:N}
 \end{figure}

 As an illustration, we plot $\scrP \, \ell^{Y_P} - 0.045 \ell^{-1}$ in Fig.\ \ref{fig:P} and $\scrN \ell^{-Y_N}$ in Fig.\ \ref{fig:N},
 both vs.\ $\ell^{-1}$, where the $\dm$ value is set at a series of values in the range $[1.1302, 1.1312]$ in steps of 0.0002,
 including the above estimate $\dm = 1.1308$.   The term $- 0.045 \ell^{-1}$ is included in Fig.\ \ref{fig:P} to remove the overall
 slope seen in the data of $\scrP$; we did not do this to the $\scrN$ data (Fig.\ \ref{fig:N}), and there the slope is evident.
 The values of $Y_P$ and $Y_N$ are obtained from Eq.\ (\ref{eq:scaling_relation}), using the exactly known values of $\beta$ and $\nu$.
 Because the leading corrections have been subtracted in Fig.\ \ref{fig:P}, it is expected that 
 the curve for the correct $\dm$ value should asymptotically become flat and reach a constant.
 Figure \ref{fig:P} shows that as $\ell$ increases, the curve for the conjectured value $217/192 \approx 1.1302$ is bending up
 while the curve for $1.1312$ is bending down.  This implies that the correct $\dm$ value should fall somewhere in between.
 A similar behavior is seen in Fig.\ \ref{fig:N}, where the curve for $1.1308$ is approximately straight 
 while those for 1.1302 and 1.1312 are bending down and up, respectively.

\section{Further simulations for 2D}
Although the conjectured number 217/192 seems to be ruled out by the data shown in Figs.\ \ref{fig:P} and \ref{fig:N}, 
a more careful analysis is still desirable. 
The above analysis makes an assumption that the leading correction is governed by $\ell^{-1}$,
but the physical origin of this term is unclear as the leading irrelevant thermal scaling field has exponent $y_i=-2$.
It is conceivable that more slowly convergent corrections exist but are not detected by the simulations up to $\ell_{\rm max}$=1024.
In particular, percolation can be regarded as a special case of biased-directed percolation with the symmetry between spatial 
and temporal directions restored \cite{Zhou12}. In this case, multiplicative and/or additive logarithmic corrections can occur in principle such that the scaling behavior of $\scrN, \scrR$, and $\scrP$ is modified as
\begin{equation}
  \scrO(\ell) \sim [\log (\ell/\ell_0)]^{y_{\rm m}} \, \ell^Y \, 
  \left(1+ 1/[\log (\ell/\ell_1)]^{y_{\rm c}} \right) \; ,
  \label{eq:log_N}
\end{equation}
where $\ell_0$ and $\ell_1$ are constants, and $y_{\rm m}$ and $y_{\rm c}$ are the associated correction exponents. 
Corrections of the $\log \log \ell$ form are also possible.
We note that due to cancellation between nominator and denominator,
the multiplicative logarithmic correction will not explicitly appear in the
ratio $Q_\scrO$ ($\scrO=\scrN, \scrR, \scrP$), for which
the scaling behavior is modified as
\begin{equation}
  Q_\scrO (\ell) = 2^Y \left(1+ 1/[\log (\ell/\ell_1)]^{y'_{\rm c}} \right) \; ,
  \label{eq:log_QN}
\end{equation}
where $y'_{\rm c}$ can be equal to $y_{\rm c}$ or $|y_{\rm m}|$, depending on the relative amplitudes
of the terms associated with them.

 To investigate this, we carried out more extensive simulations up to  $\ell_{\rm max} = 16384$.
 The number of samples was $4.5\times 10^{10}$ for $\ell \leq 1024$,  $5\times 10^9$ for 
 $1024 < \ell \leq 4096$, $10^9$ for  $4096<\ell \leq 8192$, and $3\times 10^8$ for $\ell>8192$.

\begin{figure}
 \includegraphics[scale=0.65]{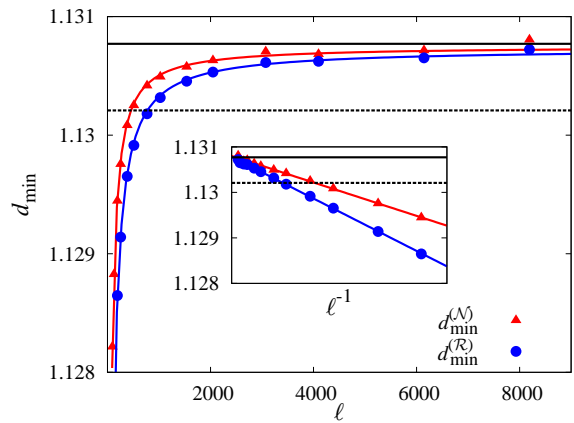}
 \centering
 \caption{(Color online) Plot of $d_{\rm min}$ versus $\ell$ in 2D, deduced from $Q_\scrN$ and $Q_\scrR$.
	 The inset shows $d_{\rm min}$ versus $\ell^{-1}$.
	 The solid and dashed horizontal lines correspond to $\dm=1.130\,77$ and $ 217/192$, respectively.
	 The red (upper) and blue (lower) curves are obtained from the fits.}
 \label{Fig:dmin_s}
 \end{figure}

 From the $Q_\scrO$ data,
 we calculate the $\dm (\ell) $ value by 
 Eqs.\ (\ref{eq:scaling_relation}) and (\ref{eq:exponent-Q}). 
 Table \ref{Tab:data_dm} displays the resulting values of $\dm (\ell)$ from the ratios $Q_\scrN$, $Q_\scrR$ and $Q_\scrP$. 
 It can be clearly seen that for $\ell \leq 3072$, the $\dm$ values that derive from $\scrN$ and $\scrR$ increase monotonically
 as $\ell$ increases.
 Further, by looking at the $\dm^{(\scrN)} (L)$ or $\dm^{(\scrR)}(L)$ data for $L = 1024, 2048, 4096L$, one can safely conclude
that the asymptotic value $\dm$ is {\it larger} than 1.1307.
 For clarity, these data are plotted in Fig.\ \ref{Fig:dmin_s}.
 The conjecture $\dm =217/192$ would mean that the monotonically increasing curves 
 for $\dm^{(\scrN)}$ and $\dm^{(\scrR)}$ must  bend downward as $\ell$ become larger, and thus a very rapid drop 
 would occur near the origin ($1/\ell \rightarrow 0$) in the inset of Fig.\ \ref{Fig:dmin_s}, which seems very unlikely.
 The $\dm^{(\scrP)}$ data are less accurate and not shown  in Fig.\ \ref{Fig:dmin_s}. 

 \begin{table}
 \begin{center}
 \begin{tabular}[t]{|l|l|l|l|}
 \hline
 $\ell $     &$\dm^{(\scrN)}$   & $\dm^{(\scrR)}$   & $\dm^{(\scrP)}$   \\
 \hline
  12      & 1.112\,909(3)\;\;& 1.102\,251(2)\;\; & 1.099\,42(3)\;\;  \\
  16      & 1.117\,007(4)    & 1.109\,204(2)     & 1.106\,44(3)      \\
  24      & 1.121\,303(4)    & 1.116\,112(2)     & 1.114\,18(3)      \\
  32      & 1.123\,540(4)    & 1.119\,588(2)     & 1.118\,10(3)      \\
  48      & 1.125\,835(4)    & 1.123\,109(2)     & 1.122\,14(3)      \\
  64      & 1.127\,007(4)    & 1.124\,902(2)     & 1.124\,20(3)      \\
  96      & 1.128\,215(4)    & 1.126\,743(2)     & 1.126\,32(3)      \\
 128      & 1.128\,826(4)    & 1.127\,685(2)     & 1.127\,38(3)      \\
 192      & 1.129\,445(4)    & 1.128\,647(2)     & 1.128\,44(3)      \\
 256      & 1.129\,755(4)    & 1.129\,140(2)     & 1.128\,99(3)      \\
 384      & 1.130\,083(4)    & 1.129\,653(2)     & 1.129\,57(3)      \\
 512      & 1.130\,251(4)    & 1.129\,914(2)     & 1.129\,83(3)      \\
 768      & 1.130\,42(2)     & 1.130\,180(6)     & 1.130\,15(9)      \\
1024      & 1.130\,49(2)     & 1.130\,317(6)     & 1.130\,36(9)      \\
1536      & 1.130\,58(2)     & 1.130\,455(7)     & 1.130\,33(9)      \\
2048      & 1.130\,63(2)     & 1.130\,532(7)     & 1.130\,27(9)      \\
3072      & 1.130\,70(3)     & 1.130\,61(2)      & 1.130\,5(2)      \\
4096      & 1.130\,68(3)     & 1.130\,62(2)      & 1.130\,6(2)      \\
6144      & 1.130\,72(7)     & 1.130\,65(4)      & 1.130\,9(5)      \\
8192\;\;  & 1.130\,80(7)     & 1.130\,72(4)      & 1.130\,8(5)      \\
\hline
\end{tabular}
\caption{Results for $\dm$ from $Q_{\scrN}$, $Q_\scrR$ and $Q_\scrP$ in 2D.}
\label{Tab:data_dm}
\end{center}
\end{table}

We fitted the $\dm (\ell) $ data by
 \begin{equation}
 \dm (\ell) = \dm + b_1 \, \ell^{y_1} + b_2 \, \ell^{-2} \ ,
 \label{eq:fit_dm}
 \end{equation}
  using a least-squares criterion.
  The data for small $\ell<\ell_{\rm min}$ were gradually excluded to
 see how the residual $\chi^2$ changes with respect to $\ell_{\rm min}$.
 Table \ref{Tab:fit_dm} lists the fitting results for $\dm^{(\scrN)}$, $\dm^{(\scrR)}$, and $\dm^{(\scrP)}$.
 From these fits, we obtain $\dm^{(\scrN)} = 1.130 \, 77(3)$, $\dm^{(\scrR)} = 1.130 \, 77(2)$ and $\dm^{(\scrP)} = 1.130\,66(15)$,
 Note that to account for potential systematic errors, the error bars in these final estimates are taken to be significantly larger
than those statistical ones in Table II.
 Considering the stability of the fit results in Table II, we believe that the estimated error margins are reliable.
 We note that the coefficient $b_2$ cannot be determined well in the fits for $\ell_{\rm min}>32$.
 Thus, fits with $b_2=0$ were also carried out, and the results agree with our above estimates of $\dm$. 

 \begin{table}
 \begin{center}
 \begin{tabular}[t]{|l|l|l|l|l|l|l|l|}
 \hline
              &$\ell_{\rm min}$&$\chi^2$& d.o.f  &  $\dm$     &   $b_1$    & $b_2$       & $y_1$   \\
 \hline
$\dm^{(\scrN)}$ &16           & 13  &  15 & 1.130\,759(5)\;& $-0.206(2)$  & 0.15(2)     & $-0.961(3)$\;   \\
                &24           & 12  &  14 & 1.130\,764(6)  & $-0.202(4)$  & 0.12(3)     & $-0.957(5) $    \\
                &32           & 12  &  13 & 1.130\,763(8)  & $-0.204(6)$  & 0.13(6)     & $-0.958(7)$     \\
                &48           & 11  &  12 & 1.130\,766(10) & $-0.20(1)$   & 0.1(2)      & $-0.95(1)$     \\
                &64           & 7   &  11 & 1.130\,780(12) & $-0.18(2)$   & 0.4(3)      & $-0.93(2)$     \\
 \hline
$\dm^{(\scrR)}$ &24           & 20  &  14 & 1.130\,776(4)  & $-0.265(2)$  & $-0.19(2)$    &$-0.918(2)$     \\
                &32           & 13  &  13 & 1.130\,771(4)  & $-0.270(3)$  & $-0.14(3)$    &$-$0.922(2)       \\
                &48           & 12  &  12 & 1.130\,768(5)  &$-$0.273(5)  &$-$0.08(7)    &$-$0.925(4)     \\
                &64           & 10  &  11 & 1.130\,772(7)  &$-$0.266(7)  &$-$0.2(2)     &$-$0.920(5)     \\
 \hline
$\dm^{(\scrP)}$ &16           & 8   &  15 & 1.130\,67(4)   &$-$0.39(2)   &  0.41(9)    &$-$0.99(1)      \\
                &24           & 8   &  14 & 1.130\,66(5)   &$-$0.40(4)   &  0.5(3)     &$-$0.99(2)        \\
                &32           & 8   &  13 & 1.130\,66(6)   &$-$0.41(5)   &  0.6(5)     &$-$0.99(3)      \\
                &48           & 8   &  11 & 1.130\,65(7)   &$-$0.4(1)    &  0.8(11)    &$-$1.00(5)      \\
                &64           & 8   &  11 & 1.130\,64(8)   &$-$0.4(2)    &  1(2)       &$-$1.01(8)      \\
\hline
\end{tabular}
\caption{Fitting results of $\dm$ in 2D, for various cutoffs $\ell_\mathrm{min}$.  ``d.o.f.'' stands for ``degrees of freedom.''} 
\label{Tab:fit_dm}
\end{center}
\end{table}

We also simulated critical site percolation on an $L \times L$ triangular lattice with periodic boundary conditions;
 this system is known to have zero amplitude of the leading irrelevant scaling field with exponent $y_i=-2$. 
 A row of lattice sites was chosen, and all the occupied sites on this row were assumed to belong to the same cluster. 
 The Leath-Alexandrowicz method was then used to grow the cluster. 
 The chemical radius $\ell$ of the completed cluster was measured. 
 From the scaling $\ell \sim L^{\dm}$, we determine $\dm = 1.130\,7(1)$, also ruling out the conjectured value.

 \begin{table}
 \begin{center}
 \begin{tabular}[t]{|l|l|l|l|}
 \hline
 $\ell $     &$\dm^{(\scrN)}$  & $\dm^{(\scrR)}$  & $\dm^{(\scrP)}$   \\
 \hline
  12      & 1.364\,7(2)\;\;& 1.358\,44(8)\;\; & 1.357\,6(4)\;\;  \\
  16      & 1.365\,4(2)    & 1.363\,14(8)     & 1.359\,7(4)      \\
  24      & 1.366\,8(2)    & 1.367\,22(8)     & 1.363\,1(4)      \\
  32      & 1.367\,9(2)    & 1.369\,01(9)     & 1.365\,6(4)      \\
  48      & 1.369\,4(2)    & 1.370\,69(9)     & 1.368\,4(4)      \\
  64      & 1.370\,4(2)    & 1.371\,52(9)     & 1.369\,9(4)      \\
  96      & 1.371\,6(2)    & 1.372\,42(9)     & 1.371\,8(4)      \\
 128      & 1.372\,3(2)    & 1.372\,94(9)     & 1.372\,8(4)      \\
 192      & 1.373\,1(2)    & 1.373\,50(9)     & 1.373\,8(4)      \\
 256      & 1.373\,5(2)    & 1.373\,78(9)     & 1.374\,4(4)      \\
 384      & 1.374\,1(2)    & 1.374\,19(9)     & 1.375\,0(4)      \\
 512      & 1.374\,4(2)    & 1.374\,49(9)     & 1.375\,3(4)      \\
 768      & 1.374\,6(2)    & 1.374\,7(2)      & 1.375\,4(5)      \\
1024\;\;  & 1.374\,6(3)    & 1.374\,7(2)      & 1.376\,0(5)      \\
\hline
\end{tabular}
\caption{Results for $\dm$ from $Q_{\scrN}$, $Q_\scrR$ and $Q_\scrP$ in 3D.}
\label{Tab:data_dm_3D}
\end{center}
\end{table}

 \begin{figure}
 \includegraphics[scale=0.65]{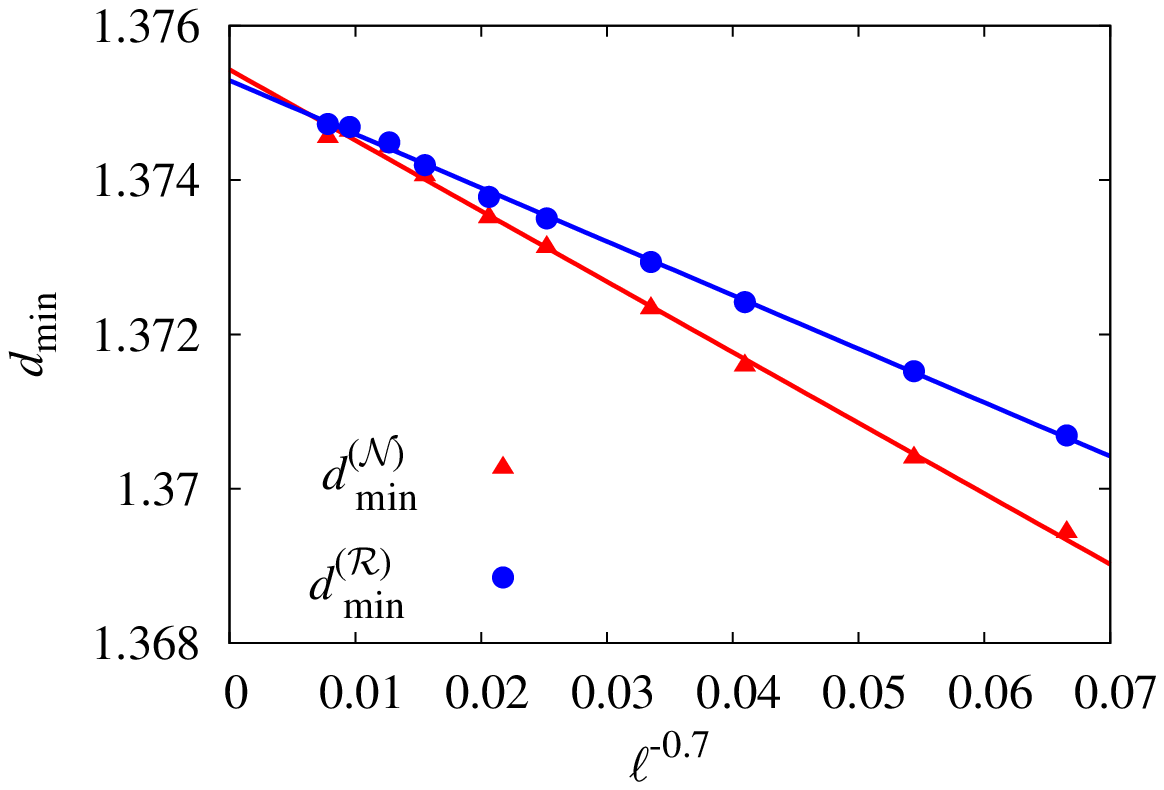}
 \centering
 \caption{(Color online) Plot of $\dm^{(\scrN)}$ and $\dm^{(\scrR)}$ versus $\ell^{-0.7}$ in 3D.}
 \label{Fig:dmin_s_3D}
 \end{figure}

 \begin{table}
 \begin{center}
 \begin{tabular}[t]{|l|l|l|l|l|l|l|l|}
 \hline
              &$\ell_{\rm min}$&$\chi^2$& d.o.f  &  $\dm$     &   $b_1$    & $b_2$       & $y_1$   \\
 \hline
$\dm^{(\scrN)}$ &8            & 6   &  11 & 1.376\,1(3)    &$-$0.062(7)  & 0.50(5)     &$-$0.57(4)      \\
                &12           & 3   &  10 & 1.375\,7(3)    &$-$0.08(2)   & 0.7(2)      &$-$0.64(6)      \\
                &16           & 2   &  9  & 1.375\,6(4)    &$-$0.09(3)   & 0.9(3)      &$-$0.68(8)      \\
                &24           & 2   &  8  & 1.375\,4(4)    &$-$0.12(6)   & 1.4(9)      &$-$0.7(2)       \\
                &32           & 2   &  7  & 1.375\,3(4)    &$-$0.2(1)    & 2(2)        &$-$0.8(2)      \\
 \hline
$\dm^{(\scrR)}$ &8            & 7   &  11 & 1.375\,4(2)    &$-$0.045(4)  &$-$1.99(3)    &$-$0.60(3)      \\
                &12           & 2   &  10 & 1.375\,7(3)    &$-$0.035(5)  &$-$1.13(7)    &$-$0.53(5)        \\
                &16           & 2   &  9  & 1.375\,8(3)    &$-$0.033(6)  &$-$1.2(2)     &$-$0.51(6)      \\
                &24           & 2   &  8  & 1.375\,7(4)    &$-$0.03(1)   &$-$1.1(3)     &$-$0.52(9)      \\
                &32           & 2   &  7  & 1.375\,7(4)    &$-$0.04(2)   &$-$0.9(6)     &$-$0.6(2)       \\
 \hline
$\dm^{(\scrP)}$ &8            & 7   &  11 & 1.376\,6(4)    &$-$0.19(4)   &  1.0(2)     &$-$0.80(6)      \\
                &12           & 4   &  10 & 1.376\,5(5)    &$-$0.23(8)   &  1.3(5)     &$-$0.8(1)         \\
                &16           & 3   &  9  & 1.376\,5(5)    &$-$0.2(1)    &  1.3(9)     &$-$0.8(2)       \\
                &24           & 3   &  8  & 1.376\,5(7)    &$-$0.2(2)    &  1(2)       &$-$0.8(3)       \\
\hline
\end{tabular}
\caption{Fitting results of $\dm$ in 3D. }
\label{Tab:fit_dm_3D}
\end{center}
\end{table}

\section{Results for 3D}
  We simulated bond percolation on the simple-cubic lattice at the central value of the recently estimated critical point
 $p=0.248\,811\,8(1)$ \cite{Junfeng}, which is slightly below the previous value of 
 $p = 0.248\,812\,6(5)$ \cite{LorenzZiff}.
 The simulation was carried up to $\ell_{\rm max} = 2048$, with the number of samples $7\times 10^9$ for $\ell\leq1024$
 and $2\times 10^9$ for $\ell > 1024$.
 Analogous to the procedure on the square lattice, we sampled 
 $\scrN$, $\scrR$, and $\scrP$ and studied the ratios $Q_{\scrN}$, $Q_{\scrR}$, and $Q_{\scrP}$.
 The values of $\dm^{(\scrO)}$ deduced from these ratios with $\beta/\nu = 0.4774(1)$
are listed in Table \ref{Tab:data_dm_3D}.
 The fitting results are shown in Table \ref{Tab:fit_dm_3D} and yield 
 $\dm^{(\scrN)} = 1.375\,6(6)$, $\dm^{(\scrR)} = 1.375\,7(6)$,
 $\dm^{(\scrP)} = 1.376\,5(10)$, and $y_1= -0.7 (2)$.
 The data of $\dm^{(\scrN)}(\ell)$ and $\dm^{(\scrR)}(\ell)$ versus $\ell^{-0.7}$ are further shown in Fig.\ \ref{Fig:dmin_s_3D},
 where the exponent $0.7$ reflects the value of $y_1$.

\section{Conclusion}
 In conclusion, we determined the shortest-path fractal dimension $\dm$ for percolation in 2D and 3D to be
$1.130\,77(2)$ and $1.375\,6(6)$, respectively.
 For the 2D value, we use the result which follows from $\scrR(\ell)$ and has the smallest error bars.
 The precision of these numbers is increased compared to the current most accurate values that we are aware of.
 The conjectured value in 2D, $\dm = 217/192$ \cite{DengEtAl10} is ruled out with a high probability.
 Grassberger’s earlier conjecture $\dm = 26/23$ \cite{Grassberger-92-0} is also ruled out.
 The 3D result represents a substantial increase in precision over the previous values of $1.34(1)$ \cite{HerrmannStanley88} and 1.374(4) \cite{Grassberger-92-1}.
 
 Y.D. is indebted Alan D. Sokal for valuable discussions.
This work was supported in part by NSFC under Grants No. 10975127 and No. 91024026, and the Chinese Academy of Science.
R.M.Z. acknowledges support from NSF Grant No.\ DMS-0553487.
The simulations were carried out on the NYU-ITS cluster, which is partly supported by NSF Grant No. PHY-0424082.


\begin{thebibliography}{100}
\bibitem{Stauffer-1994} D. Stauffer and A. Aharony, {\it Introduction to Percolation Theory},
	2nd ed. (Taylor and Francis, London, 1994).
\bibitem{Bunde-1996} A. Bunde and S. Havlin, in {\it Fractals and Disordered Systems},
	2nd ed., edited by A. Bunde and S. Havlin (Springer, New York, 1996).
\bibitem{Grassberger-92-0} P. Grassberger, J. Phys. A {\bf 25}, 5475 (1992).
\bibitem{Grassberger-92-1} P. Grassberger, J. Phys. A {\bf 25}, 5867 (1992).
\bibitem{Grassberger-83} P. Grassberger, Math. Biosciences {\bf 63}, 157 (1983).
\bibitem{Alexandrowicz-80} Z. Alexandrowicz, Phys. Lett. A {\bf 80}, 284 (1980).
\bibitem{Pike-81} R. Pike and H. E. Stanley, J. Phys. A {\bf 14}, L169 (1981).
\bibitem{Herrmann-84} H. J. Herrmann, D. C. Hong and H. E. Stanley, J. Phys. A {\bf 17}, L261 (1984).
\bibitem{Middlemiss-80} K. M. Middlemiss, S. G. Whittington, and D. S. Gaunt, J. Phys. A {\bf 13}, 1835 (1980).
\bibitem{HavlinNossal84} S. Havlin and R. Nossal, J. Phys. A {\bf 17}, L427 (1984).
\bibitem{VannimenusNadalMartin84}
J. Vannimenus, J. P. Nadal, and H. Martin, J. Phys. A {\bf 17}, L351 (1984).
\bibitem{RammalEtAl84} R. Rammal, J. C. Angles d'Auriac, and A. Benoit, J. Phys. A {\bf 17}, L491 (1984).
\bibitem{Grassberger85} P. Grassberger, J. Phys. A  {\bf 18}, L215 (1985).
\bibitem{HerrmannStanley88} H. J. Herrmann and H. E. Stanley, J. Phys. A {\bf 21}, L829 (1988).
\bibitem{Grassberger99} P. Grassberger, J. Phys. A {\bf 32}, 6233 (1999).
\bibitem{DengEtAl10}
Y. Deng, W. Zhang, T. M. Garoni, A. D. Sokal, and A. Sportiello, Phys. Rev. E  {\bf 81}, 020102(R) (2010).
\bibitem{Larsson} T. A. Larsson, J. Phys. A {\bf 20}, L291 (1987).
\bibitem{DengBloteNienhuis04} Y. Deng, H. W. J. Bl\"ote, B. Nienhuis, Phys. Rev. E {\bf 69}, 026114 (2004).
\bibitem{Grassberger99b} P. Grassberger, Physica A {\bf 262}, 251 (1999).
\bibitem{JacobsenZinnJustin02} J. L. Jacobsen and P. Zinn-Justin, Phys. Rev E {\bf 66}, 055102 (R) (2002).
\bibitem{TzschichholzBundeHavlin89}F. Tzschichholz, A. Bunde, and S. Havlin, Phys. Rev. A {\bf 39}, 5470 (1989).
\bibitem{Porto} M. Porto, S. Havlin, H. E. Roman, and A. Bunde, Phys. Rev. E {\bf 58}, 5205(R) (1998).
\bibitem{Ziff99} R. M. Ziff, J. Phys. A {\bf 32}, L457 (1999).
\bibitem{Zhou12}
Z. Zhou, J. Yang, R. M. Ziff and Y. Deng, Phys. Rev. E {\bf 86}, 021102 (2012).
\bibitem{Leath} P. L. Leath, Phys. Rev. B {\bf 14}, 5046 (1976).
\bibitem{GrassbergerDeLaTorre}  P. Grassberger and A. de la Torre, Annals Phys. (N.Y.) {\bf 122}, 373 (1979).
\bibitem{Cardy}
J. L. Cardy, Nucl. Phys. B {\bf 240}, 514 (1984).
\bibitem{Smirnov}
S. Smirnov, W. Werner, Math. Research Lett. {\bf 8}, 729 (2001).
\bibitem{Lawler}
G. F. Lawler, O. Schramm, W. Werner, Electron. J. Probab. {\bf 7}, 2 (2002).
\bibitem{Kesten-1987}
H. Kesten, Comm. Math. Phys. {\bf 109}, 109 (1987).
\bibitem{Deng-05}
Y. Deng and H. W. J. Bl\"ote, Phys. Rev. E {\bf 72}, 016126 (2005).
\bibitem{Martins-03-PRE}
P. H. L. Martins and J. A. Plascak, Phys. Rev. E {\bf 67}, 046119 (2003).
\bibitem{Tomita-JPS-02}
Y. Tomita and Y. Okabe, J. Phys. Soc. Jpn. {\bf 71}, 1570 (2002).
\bibitem{Ballesteros-99-JPA}
H. G. Ballesteros, L. A. Fern\'{a}ndez, V. Mart\'{i}n-Mayor, A. Mu\~{n}oz Sudupe, G. Parisi, and
J. J. Ruiz-Lorenzo, J. Phys. A {\bf 32}, 1 (1999).
\bibitem{Junfeng}
J. Wang, Z. Zhou, and Y. Deng (unpublished).
\bibitem{LorenzZiff}
C. D. Lorenz and R. M. Ziff, Phys. Rev. E {\bf 57}, 230 (1998).
\end{thebibliography}
\end{document}